\documentclass[aps,prb,superscriptaddress,nofootinbib,longbibliography]{revtex4-2}

\usepackage{graphicx}
\usepackage{bm}
\usepackage{amsmath}
\usepackage{amssymb}
\usepackage{amsthm}
\usepackage{hyperref}
\providecommand{\doi}[1]{\href{https://doi.org/#1}{doi:#1}}

\newtheorem{theorem}{Theorem}
\newtheorem{lemma}{Lemma}
\newtheorem{proposition}{Proposition}

\newcommand{\BZ}{\mathbb{T}}
\newcommand{\Om}{\Omega}
\newcommand{\dd}{\mathrm{d}}
\newcommand{\vol}{\,\mathrm{vol}}

\newcommand{\A}{\mathcal{A}}
\newcommand{\B}{\mathcal{B}}
\newcommand{\Var}{\mathrm{Var}}
\newcommand{\Cov}{\mathrm{Cov}}

\begin{document}

\title{The Hodge structure of Berry-phase transport:\\
topology, geometry, and noise}

\author{Zhi-Wei Wang}
\affiliation{College of Physics, Jilin University, Changchun 130012, China}
\affiliation{Department of Computer Science, University of York,
York YO10 5GH, United Kingdom}

\author{Samuel L. Braunstein}
\email{sam.braunstein@york.ac.uk}
\affiliation{Department of Computer Science, University of York,
York YO10 5GH, United Kingdom}

\begin{abstract}
We show that the Hodge--de Rham decomposition of the Berry curvature organises
the transport of a Bloch band \emph{and its fluctuations} within a single
geometric structure. Splitting the curvature into $L^2$-orthogonal harmonic,
exact, and co-exact sectors yields a dictionary for both moments of the
current. In the mean response the harmonic sector carries the topological
anomalous-Hall conductivity, the exact sector the Fermi-surface geometry and
the antisymmetric Berry-curvature dipole of polar metals, and, in three
dimensions, the co-exact sector the chiral anomaly, quantised by the Weyl-node
charges. In the fluctuations, described by a particle-conserving stochastic
Boltzmann equation constrained by the fluctuation-dissipation theorem (FDT),
the harmonic sector is silent, so topological transport is noiseless, while the
field-driven noise is sourced by the geometric sectors (solely the exact sector
in two dimensions); current-noise spectroscopy therefore separates global band
topology from local band geometry. We prove that the harmonic null-space
protection is dimension-independent, and we settle the remaining sector: the
co-exact (monopole) sector carries no conservation law and, under the thermal
sampling measure, mixes with the exact sector at $\mathcal{O}(1)$, so it
furnishes no clean noise observable. The separation the noise provides is
therefore two-way, topology versus geometry, in both two and three dimensions.
\end{abstract}

\maketitle

\section{Introduction}

Berry-curvature transport spans two regimes that sit awkwardly together. The
anomalous Hall effect (AHE) is reactive, quantised, and independent of
scattering, the textbook example of dissipationless
transport~\cite{tknn1982,haldane2004}; the nonlinear Hall effect driven by the
Berry-curvature dipole is dissipative and scales with the relaxation
time~\cite{sodemann2015,ma2019}. Both are transverse currents built from the
same object, the Berry curvature of the occupied bands, and the tension between
them is usually handled case by case. The semiclassical theory expresses a wide
class of response coefficients as integrals of the Berry curvature over the
Brillouin zone~\cite{sundaram1999,xiao2010}, computed component by component in
a fixed gauge, which obscures their global structure and is numerically delicate
near band-touching points.

We take the Berry curvature as what it is, a differential two-form on the
Brillouin torus, and apply the Hodge--de Rham decomposition. This splits it into
three orthogonal pieces: a harmonic sector, which carries the topological
content; an exact sector, which carries the Fermi-surface geometry; and, in
three dimensions, a co-exact sector, sourced by the Weyl nodes. The
decomposition is coordinate-free and the sectors are $L^2$-orthogonal, so any
linear functional of the curvature splits cleanly along them. Which sectors are
present is fixed by dimension: in two dimensions the curvature is a top-degree
form with an empty co-exact sector, while in three dimensions the co-exact
sector is non-empty precisely when Weyl nodes are present.

The central claim of this paper is that the same three sectors govern not one
but \emph{both} moments of the current. For the mean current the dictionary is:
harmonic gives the topological anomalous-Hall conductivity; exact gives the
Fermi-surface geometry and the antisymmetric Berry-curvature dipole, a
symmetry-protected nonlinear-Hall observable nonzero only in polar metals; and,
in three dimensions, co-exact gives the chiral anomaly as its projection,
quantised by node charge and carrying a gyrotropic-optical signature. For the
fluctuations the dictionary is sharper still. Placing the band in a
particle-conserving stochastic Boltzmann equation whose noise is fixed by the
fluctuation-dissipation theorem (FDT), we find that the harmonic sector is
silent: the topological flux sits in the null space of the macroscopic noise
operator, coupling only to the conserved total particle number, so a topological
current carries no thermal noise, and Haldane's split of the anomalous Hall
response follows without recourse to Dirac strings. In two dimensions the
field-driven current noise is sourced entirely by the exact geometric sector;
in three dimensions it is carried by the two geometric sectors, exact and
co-exact, jointly. This result is
complementary to the intrinsic, quantum-metric noise identified recently by
Bhowmick and Agarwal~\cite{BA26}, and to the observation of
Neupert, Chamon, and Mudry that equilibrium current fluctuations encode the
quantum geometric tensor~\cite{NCM13}; the noise here is extrinsic,
scattering-time dependent, and intraband.

The consequence is an experiment. In the mean current, topology and geometry are
superposed and hard to disentangle; in the noise they separate, because one is
silent and the other is the source. Current-noise spectroscopy therefore
isolates local band geometry from global band topology, a separation that the
mean response cannot provide. The deliverable of the paper is this unified
sector-to-observable dictionary and the separation principle that follows from
it.

We are explicit about the reach of what is established, and we resolve here
the one question that the two-moment logic makes conspicuous: the noise signature
of the co-exact (monopole) sector, which exists only in three dimensions. The
mean-current dictionary holds in two and three dimensions. The noise results, the
harmonic null space and the exact-sector source, were first established for the
two-dimensional Brillouin zone. We lift them to three dimensions and prove that
the harmonic null-space protection is dimension-independent: the topological
anomalous-Hall vector remains rigorously noiseless. We then compute the co-exact
noise directly and show, as a theorem with numerical corroboration, that it is
\emph{not} a clean sector observable. The reason is structural: the protection
that silences the harmonic sector is a conservation law, and the co-exact sector
has none, so under the thermal sampling measure it mixes with the exact sector at
$\mathcal{O}(1)$. The dictionary's last cell is thereby filled with a definite
no-go, and the clean statement that survives, in both dimensions, is the two-way
separation of topology from geometry.

\section{The Hodge decomposition of Berry-phase response}

Let the Brillouin zone be the flat torus $\BZ^d$ and let $\Om$ be the Berry
curvature $2$-form of an isolated band or group of bands. On a closed oriented
Riemannian manifold the Hodge theorem gives the unique orthogonal
decomposition~\cite{nakahara2003,frankel2011}
\begin{equation}\label{eq:hodge}
\Om = \underbrace{\gamma}_{\text{harmonic}}
    + \underbrace{\dd\A}_{\text{exact}}
    + \underbrace{\delta\B}_{\text{co-exact}},
\end{equation}
where $\gamma$ satisfies $\Delta\gamma = 0$ and represents the cohomology class
$[\Om] \in H^2(\BZ^d;\mathbb{R})$, $\A$ is a $1$-form, $\B$ is a $3$-form, and
$\dd,\delta$ are the exterior derivative and co-derivative. The three summands
are mutually orthogonal in the $L^2$ inner product
$\langle\alpha,\beta\rangle = \int_{\BZ^d}\alpha\wedge\star\beta$.

Any Berry-phase linear-response coefficient of the form
\begin{equation}\label{eq:response}
\sigma = \int_{\BZ^d} f\,\Om,
\end{equation}
with $f$ a smooth weight (typically the occupation function), inherits the
decomposition~\eqref{eq:hodge}. We now state the physical identification of the
three sectors, which is the organising claim of the paper.

\paragraph{Harmonic sector (topological).}
The harmonic representative $\gamma$ is fixed by the cohomology class of $\Om$
and is therefore rigid under any smooth deformation of the band structure that
does not change the topology. In two dimensions $\gamma$ is the constant
$2$-form whose integral is $2\pi c_1$, and it carries the quantised anomalous
Hall conductivity. In three dimensions $\gamma$ encodes the anomalous Hall vector
set by the separation of the Weyl nodes (modulo a reciprocal-lattice vector). In
both cases $\gamma$ is a \emph{constant} form; this single fact, we show below,
is the reason the topological sector is noiseless.

\paragraph{Exact sector (geometric).}
The exact part $\dd\A$ is fixed by the geometry of the occupied states. It
carries the Berry curvature dipole and its higher-order analogues, and its
contribution to any response coefficient of the form~\eqref{eq:response} vanishes
identically for a completely filled band, where $f$ is constant and
$\int f\,\dd\A = 0$; for partial filling it reduces, after integration by parts,
to a Fermi-surface integral.

\paragraph{Co-exact sector (monopole).}
The co-exact part $\delta\B$ is sourced by the failure of $\Om$ to be closed:
\begin{equation}\label{eq:source}
\Delta\B = \dd\Om = 2\pi\sum_n \chi_n\,\delta^{3}(\bm{k}-\bm{k}_n),
\end{equation}
where $\bm{k}_n$ are the Weyl nodes and $\chi_n = \pm 1$ their chiralities. This
sector is identically zero in two dimensions, where $\Om$ is top-degree and
$\dd\Om = 0$ automatically. It first appears in three dimensions. Equation
\eqref{eq:source} is solvable on the compact torus only if its right-hand side
integrates to zero,
\begin{equation}\label{eq:nn}
\sum_n \chi_n = 0,
\end{equation}
the Nielsen--Ninomiya theorem, here appearing as the Hodge-theoretic solvability
condition for the co-exact sector. When it holds, the solution is
$\B(\bm{k}) = 2\pi\sum_n\chi_n\,G(\bm{k},\bm{k}_n)$, with $G$ the Green's function
of the Laplacian on $\BZ^3$.

\paragraph{Vector (Helmholtz) form.}
Identifying the $2$-form with the pseudovector field $\bm{\Om}$
($\Om_a = \tfrac12\epsilon_{abc}\Om_{bc}$), Eq.~\eqref{eq:hodge} is the
Helmholtz decomposition
$\bm{\Om} = \langle\bm{\Om}\rangle + \bm{\Om}_\perp + \bm{\Om}_\parallel$ into a
constant field (harmonic), a transverse divergence-free field (exact), and a
longitudinal curl-free field (co-exact). The Helmholtz projection assigns the
entire divergence to the co-exact field, while the exact and harmonic fields are
divergence-free; this is the representation used in all our numerics.

\section{Mean transport: sectors as transport coefficients}

\subsection{Two dimensions: topological and geometric sectors}

In two dimensions the decomposition has only two nontrivial pieces. The Berry
curvature is the top-degree form
$\Om = \Om_{xy}(\bm{k})\,\dd k_x\wedge\dd k_y$, which is closed, so
$\delta\B = 0$ identically. The harmonic part is the constant form
$\gamma = \bar{\Om}\,\dd k_x\wedge\dd k_y$ with
$\bar{\Om} = (2\pi)^{-2}\int_{\BZ^2}\Om_{xy}\vol$, and the exact part is the
remainder $\dd\A = \Om - \gamma$, which has zero Brillouin-zone integral by
construction.

\paragraph{Anomalous Hall effect.}
The anomalous Hall conductivity
$\sigma_{xy} = -(e^2/\hbar)(2\pi)^{-2}\int_{\BZ^2} f\,\Om$ splits accordingly.
For a filled band ($f=1$) only the harmonic part survives and yields the
quantised value $\sigma_{xy} = -(e^2/h)\,c_1$~\cite{tknn1982}. For a partially
filled band the harmonic sector contributes the filling-weighted term
$-(e^2/h)\,c_1\bar{f}$, no longer quantised, while the exact sector carries the
entire remainder; integrating the latter by parts turns it into a Fermi-surface
integral of the Berry potential, in the spirit of Haldane's formulation of the
AHE as a Fermi-liquid property~\cite{haldane2004}. The topological and geometric
contributions are separated cleanly and gauge-invariantly, without introducing a
Dirac string.

\paragraph{Nonlinear Hall effect.}
The nonlinear Hall response of a time-reversal-invariant band is governed by the
Berry curvature dipole
$D_a = (2\pi)^{-2}\int_{\BZ^2} f\,\partial_a\Om_{xy}\,\dd^2k$. Because
$\partial_a\Om_{xy}$ has zero Brillouin-zone average, the dipole lives entirely
in the exact sector: its harmonic component vanishes identically. This reproduces
the Sodemann--Fu result~\cite{sodemann2015,ma2019} and makes manifest why the
nonlinear Hall response is a purely geometric (Fermi-surface) quantity with no
topological contribution. We have verified numerically, in the two-band
Qi--Wu--Zhang model, that the Berry curvature computed by the gauge-invariant
Fukui--Hatsugai--Suzuki plaquette method~\cite{fhs2005} has its full quantised
Hall number carried by the harmonic sector, with the exact remainder integrating
to zero to machine precision.

\subsection{Three dimensions: the co-exact sector and the chiral anomaly}

In three dimensions the Berry curvature $2$-form on $\BZ^3$ is no longer
top-degree, and it need not be closed. Weyl nodes are momentum-space monopoles at
which $\dd\Om$ is a delta-function source, Eq.~\eqref{eq:source}. The Hodge
decomposition then has a nonzero co-exact sector $\delta\B$, absent in two
dimensions. Since the curvature of a gapless band is singular at the nodes
($|\Om| \sim 1/r^2$, not square-integrable in three dimensions), the
decomposition is understood in the distributional sense; every statement below
either concerns mode-by-mode (spectral) projections, which are well defined, or
is formulated at finite lattice resolution.

\paragraph{Lattice Weyl model.}
We use the minimal two-band Weyl semimetal
\begin{align}\label{eq:weyl}
H(\bm{k}) ={}& t\sin k_x\,\sigma_x + t\sin k_y\,\sigma_y \nonumber\\
&+ \bigl[m(2-\cos k_x-\cos k_y) + 2t_z(\cos k_z-\cos k_0)\bigr]\sigma_z,
\end{align}
with Weyl nodes at $\bm{k} = (0,0,\pm k_0)$ of opposite chirality. We
use $t = t_z = 1$, $m = 2$, $k_0 = \pi/2$, for which the two-node
condition is satisfied with a margin. In the mean-transport
verifications the Berry curvature is computed by the gauge-invariant
Fukui--Hatsugai--Suzuki plaquette method~\cite{fhs2005}; in the
fluctuation calculations of Sec.~4 we use instead the closed-form
curvature of the two-band Hamiltonian $H=\bm{d}\cdot\bm{\sigma}$, which
is exact. The Hodge decomposition of $\Om$ is obtained as the
Helmholtz--Hodge decomposition of the associated Berry-curvature vector
field, computed by fast Fourier transform on grids of odd linear size
$N$. (The archived verification and figure scripts sample on
half-offset grids, $k_i=(l+\tfrac12)\,2\pi/N$, which keep every node off
the grid points independently of the parity of $N$; odd $N$ remains
preferable because it excludes the Nyquist mode.)
Odd $N$ (together with $k_0=\pi/2$) ensures that no grid
\emph{point} coincides with a node and, by excluding the Nyquist mode,
that the Fourier projectors onto the three sectors are orthogonal to
machine precision; for even $N$ the residual exact/co-exact overlap is
at the $10^{-2}$ level. The reconstructed curvature matches the input to
a relative $L^2$ error of $\sim\!10^{-16}$, and the monopole charges,
evaluated by Gauss's law as the difference of the Berry flux through two
lattice planes flanking each node (the region between two $k_z$ planes
of the torus is closed), are $\pm1$ in units of $2\pi$ to six digits and
sum to zero, Eq.~\eqref{eq:nn}. The slice Chern number $C(k_z) =
(2\pi)^{-1}\int\Om_{xy}\,\dd k_x\,\dd k_y$ is integer-quantised on every
slice except the two immediately adjacent to each node, where the
discretised singular field is not yet resolved; for the
regularisation~\eqref{eq:weyl} with $m=2$ the Chern-nontrivial slices
are those with $|k_z|>k_0$, i.e.\ the unit plateau connects the two
nodes through the zone boundary, consistent with the anomalous-Hall
vector being the node separation modulo a reciprocal-lattice vector;
only its $k_z$-average
$\bar{C} = 1 - k_0/\pi$ is the harmonic sector [whose finite-$N$ deficit
closes slowly, as $\mathcal{O}(1/N)$, another near-node discretisation
effect], while the quantised steps of $C(k_z)$ across the nodes are
carried by the \emph{co-exact} sector, since the flux of the exact
(curl) part through any closed slice vanishes while the co-exact
(gradient) part jumps by $2\pi\chi_n$ at each monopole. This last point
matters for the noise analysis below: in three dimensions the quantised
topological jumps do not live in the harmonic sector, and are therefore
not automatically protected by the null-space mechanism that silences
it.

\begin{figure}[t]
\centering
\includegraphics[width=\textwidth]{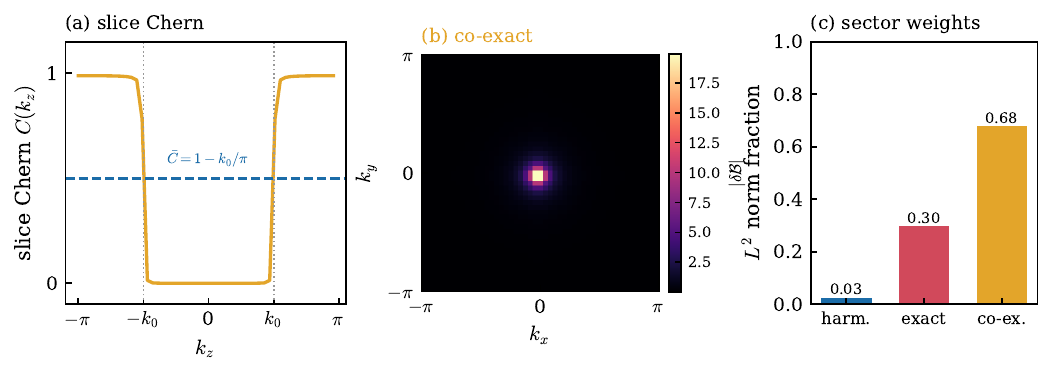}
\caption{\label{fig:hodge}Hodge decomposition of the Berry
curvature in the lattice Weyl model, Eq.~\eqref{eq:weyl}, with
$t=t_z=1$, $m=2$, $k_0=\pi/2$, on a half-offset Fourier grid of odd
size $N=49$, so that no node lies on a lattice point. (a)~The slice
Chern number $C(k_z)=(2\pi)^{-1}\int\Om_{xy}\,\dd k_x\,\dd k_y$ (solid)
steps by $\chi_n$ at each Weyl node; for the regularisation
\eqref{eq:weyl} the unit plateau occupies $|k_z|>k_0$, the node
separation taken through the zone boundary. Its $k_z$-average $\bar
C=1-k_0/\pi$ (dashed) is the harmonic sector, while the quantised steps
are carried by the co-exact sector, because the exact (curl) part has
zero flux through any closed slice. (b)~The co-exact field magnitude
$|\delta\B|$ on the $k_z$ slice adjacent to the node at $+k_0$ (node
centred), concentrated at the node, whose low-$|\bm G|$ weight
reproduces the point-monopole prediction of Eq.~\eqref{eq:normsum}.
(c)~The $L^2$ weight of $\Om$ carried by each Hodge sector at this
resolution (harmonic $3\%$, exact $30\%$, co-exact $68\%$): the co-exact
sector carries the largest share, and the exact/co-exact split is
non-universal, being model- and resolution-dependent because the
near-node curvature is not square-integrable, while the harmonic
component itself is fixed by the topology.
This is the mean-transport structure the noise analysis of Sec.~4
inherits: the topological content of the slice-Chern jumps sits in the
co-exact, not the harmonic, sector, which is why Theorem~\ref{thm:3d}
does not protect it.}
\end{figure}

\paragraph{Universal result I: the chiral anomaly is the co-exact sector.}
The chiral-anomaly response, the current induced by parallel $\bm{E}$
and $\bm{B}$~\cite{armitage2018,fukushima2008}, is the projection of the
Berry-phase response onto the co-exact sector. The anomaly coupling to a
uniform drive with occupation weight $W$ is $A = \int_{\BZ^3}
W\,(\dd\Om)$; using $\dd\Om = \dd(\delta\B)$ and integrating by parts,
\begin{equation}\label{eq:proj}
A = \int_{\BZ^3} W\,\dd(\delta\B) = -\int_{\BZ^3} (\dd W)\wedge\delta\B,
\end{equation}
so $A$ couples to the co-exact sector alone: the harmonic sector, being
constant, has zero exterior derivative, and the exact sector contributes
nothing because $\dd(\dd\A)=0$. Because $\dd\Om$ is the delta-function
monopole source, the coefficient reduces to
\begin{equation}\label{eq:quant}
A = \sum_n q_n\,W(\bm{k}_n), \qquad q_n = 2\pi\chi_n,
\end{equation}
quantised by the node charges. In the lattice model, with $W = \sin
k_z$, the directly computed coupling extrapolates to the quantised node
sum with the deficit vanishing quadratically in the lattice spacing.
This co-exact identification is directly measurable as the gyrotropic
magnetic response / the natural optical activity of a Weyl
metal~\cite{zhong2016,mapesin2015}, which couples to the same monopole
source with an occupation weight set by the band energy at the nodes.
Equation~\eqref{eq:proj} is the sharp statement that the chiral-anomaly
response and the anomalous Hall response are orthogonal: the former is
the co-exact sector, the latter the harmonic sector.

\paragraph{Universal result II: a lower bound on the co-exact sector.}
The magnitude of $\delta\B$ is model-dependent, but its infrared weight is
bounded below by the topological node data. Working with the spectrally truncated
norm $\|\delta\B\|^2_{\leq\Lambda}$ (the contribution of reciprocal-lattice modes
$0 < |\bm{G}| \leq \Lambda$, a lower bound on the full norm of any ultraviolet
regularisation),
\begin{equation}\label{eq:normsum}
\|\delta\B\|^2_{\leq\Lambda} = \frac{(2\pi)^2}{V}
\sum_{0<|\bm{G}|\leq\Lambda}\frac{|S(\bm{G})|^2}{|\bm{G}|^2},
\quad S(\bm{G}) = \sum_n \chi_n\,e^{-i\bm{G}\cdot\bm{k}_n},
\end{equation}
with $V=(2\pi)^3$ the volume of the Brillouin torus and each term non-negative. For a minimal Weyl pair the truncated sum cannot vanish
for any $\Lambda \geq 1$ unless the nodes physically merge, so
\begin{equation}\label{eq:bound}
\|\delta\B\|^2 \;\geq\; \|\delta\B\|^2_{\leq\Lambda}
\;\geq\; c_\Lambda(\Delta k)\,\sum_n \chi_n^2,
\end{equation}
with $c_\Lambda(\Delta k)>0$ depending only on the node separation, the cutoff,
and the torus. As the opposite-chirality nodes approach and annihilate,
$|S(\bm{G})|^2 = 2(1-\cos\bm{G}\cdot\Delta\bm{k}) = \mathcal{O}(\Delta k^2)$, and
the bound correctly degrades to zero. The infrared weight of the co-exact sector
for a minimal Weyl semimetal cannot be made arbitrarily small so long as the
nodes are separated and charged. We have checked the spectral form
\eqref{eq:normsum} directly against the lattice model: the low-$|\bm G|$ weight
of the numerically computed co-exact field reproduces the point-monopole
prediction $\sum_{0<|\bm{G}|\leq\Lambda}|S(\bm G)|^2/|\bm G|^2$ to better than
$1\%$ for every $\Lambda\leq 6$ (e.g.\ $\|\delta\B\|^2_{\leq 2}=5.513$ measured
versus $5.517$ predicted; $\|\delta\B\|^2_{\leq 6}=21.43$ versus $21.57$),
confirming that the infrared tail of the co-exact sector is the two-point
monopole structure and not a lattice artefact.

\subsection{The exact sector as a measurable dipole, and the smooth gauge}

The exact sector is measurable too. For any scalar $g$ the functional
$\int_{\BZ^3}g\,(\nabla\times\bm{\Om})$ is exact-sector-only, because the curl
annihilates the harmonic (constant) and co-exact (gradient) parts identically.
Taking $g=f_0$ gives the observable
\begin{equation}\label{eq:Pexact}
\bm{P}(\mu)=\int_{\BZ^3}f_0\,(\nabla\times\bm{\Om})
=\oint_{\mathrm{FS}}\hat{\bm{n}}\times\bm{\Om}\,\dd S,
\end{equation}
the circulation of the Berry curvature around the Fermi surface, which is
precisely the antisymmetric part of the Berry-curvature dipole,
$P_l=\epsilon_{lij}D_{ij}$. Under a point-group operation $\bm{P}$ transforms as
a \emph{polar} vector, so by Neumann's principle it is nonzero only in the ten
polar (pyroelectric) classes; the exact-sector response lives specifically in
polar metals such as $T_d$-WTe$_2$ or
$T_d$-TaIrTe$_4$~\cite{ma2019,kumar2021}, where it is extracted as the sum
$\chi_{xxz}+\chi_{yyz}$ of the $\tau^1$ Berry-dipole conductivities~\cite{nye1985}.
Finally, the exact sector $\dd\A$ provides a globally smooth, gauge-invariant
proxy for the Berry curvature free of Dirac-string singularities; in the
Coulomb--Hodge gauge it minimises the momentum-space spread and, for a single
trivial band, coincides with the maximally localised Wannier
gauge~\cite{marzari2012}.

\medskip
The mean-transport dictionary is thus complete in both dimensions: harmonic
$\leftrightarrow$ anomalous Hall, exact $\leftrightarrow$ nonlinear-Hall dipole,
co-exact $\leftrightarrow$ chiral anomaly.

\section{Fluctuations: the same sectors control the noise}

We now show that the same decomposition organises the second moment of the
current. The deterministic semiclassical Boltzmann equation omits the thermal
noise demanded by the fluctuation-dissipation theorem~\cite{Kubo66,LL57,Braun26}.
Assuming a spatially homogeneous macroscopic system, the particle-conserving
Langevin--Boltzmann equation is
\begin{equation} \label{eq:stoch_boltz}
\partial_t f + \dot{\bm{k}} \cdot \nabla_{\bm{k}} f
= \mathcal{C}[f] + \xi(\bm{k}, t),
\end{equation}
with $\mathcal{C}[f]$ a density-conserving collision operator
($\int \mathcal{C}[f]\,\dd^dk = 0$). To maintain the equilibrium fluctuations of
a canonical Fermi gas, where macroscopic charge neutrality enforces exact
particle conservation ($\int \xi\,\dd^dk \equiv 0$), the FDT fixes the noise
correlator
\begin{equation}\label{eq:fdt_noise}
\langle \xi(\bm{k},t)\xi(\bm{k}',t')\rangle
= \frac{2}{\tau}\frac{(2\pi)^d}{V_{\mathrm{sys}}}
\Big[ D(\bm{k})\delta(\bm{k}-\bm{k}')
- \frac{D(\bm{k})D(\bm{k}')}{\mathcal{Z}} \Big]\delta(t-t'),
\end{equation}
with $D(\bm{k}) = f_0(1-f_0) = \tfrac14\,\mathrm{sech}^2[\beta(\varepsilon-\mu)/2]$
and $\mathcal{Z} = \int_{\BZ} D\,\dd^dk$. On the compact torus $D(\bm{k})$ is
strictly bounded above zero at any $T>0$ (Appendix~\ref{app:A}), so the Fokker--Planck
operator is uniformly elliptic on the physical subspace of internal momentum
exchanges; the negative cross-correlation removes exactly one mode, the total
particle number, whose eigenvalue is zero.

\subsection{The null-space theorem: topology is noiseless (2D)}

Substituting the Hodge decomposition
$\Om = 2\pi c_1/A_\BZ\,\vol + \dd\A$ into the anomalous Hall current partitions
the mean response into a filling-weighted TKNN term and a Fermi-surface term,
recovering Haldane's split without Dirac strings, exactly as in the mean-transport
section. The new content is dynamical. The time-dependent fluctuation of the Hall
current is
\begin{equation}\label{eq:noise_split}
\delta J_{\mathrm{AHE}}^i(t)
= -e^2\epsilon^{ij}E_j\frac{1}{(2\pi)^2}
\Big[ \frac{2\pi c_1}{A_\BZ}\!\int_{\BZ}\!\delta f\,\vol
+ \int_{\BZ}\!\delta f\,\dd\A \Big].
\end{equation}
The first term is proportional to the density fluctuation
$\delta n(t)=\int_{\BZ}\delta f\,\vol \propto \delta N(t)$, which the
particle-conserving noise pins to zero, $\delta N(t)\equiv 0$. Hence the
harmonic--harmonic and harmonic--exact channels of the current-current
correlator vanish identically, leaving the exact--exact channel as the sole
source of noise.

\begin{theorem}[Decoupling from thermal fluctuations]\label{thm:fdt}
Under particle-conserving stochastic dynamics with the noise correlator
\eqref{eq:fdt_noise}, the non-equilibrium current-current correlator of the
harmonic sector vanishes identically for all $t,t'$. The non-equilibrium excess
noise power is $S_{\mathrm{AHE}}(0) \propto \Var_p(\Om_{\mathrm{geom}})$,
independent of $c_1$.
\end{theorem}

Explicitly (Appendix~\ref{app:B}), the low-frequency excess noise is
\begin{equation}\label{eq:variance}
S_{\mathrm{AHE}}(0) = \Big(\frac{e^2 E_x}{(2\pi)^2}\Big)^2
\frac{2\tau(2\pi)^2}{V_{\mathrm{sys}}}\,\mathcal{Z}\;
\Var_{p}(\Om_z),
\end{equation}
with $\Var_p$ the variance under the thermal envelope
$p(\bm{k})=D(\bm{k})/\mathcal{Z}$. Because the harmonic sector is a global
constant, $\Var_p(2\pi c_1/A_\BZ + \Om_{\mathrm{geom}})=\Var_p(\Om_{\mathrm{geom}})$:
the Chern number cancels exactly. Field-driven noise is sourced exclusively by
the exact geometric sector. This yields three sharp predictions: the Chern number
is a noiseless quantum number; the $\mathcal{O}(E^2)$ transverse noise scales out
the topological response, giving a direct noise-spectroscopy probe of band
geometry $\dd\A$; and in the QAH limit $T\to0$ the gapped thermal envelope
collapses ($\mathcal{Z}\to0$ exponentially, and the prefactor
$\mathcal{Z}\Var_p\to0$ with it) so the bulk excess noise vanishes, preserving
exact quantisation while obeying the FDT. A Corbino geometry with inductive ($\delta
B_z$) readout isolates the azimuthal Hall noise from radial Ohmic noise; the
Arrhenius activation of the quantisation deviations observed by Bestwick
\emph{et al.}~\cite{GG15} in Cr-doped (Bi,Sb)$_2$Te$_3$ follows the same
$D=f_0(1-f_0)$ envelope that controls Eq.~\eqref{eq:variance}.

\paragraph{Scope: bulk noise, not edge transport.}
The noiselessness of the harmonic sector is a statement about the bulk
semiclassical response. Theorem~\ref{thm:fdt} concerns the thermal
fluctuation of the bulk Berry-curvature current, not the edge channels
that carry the quantised conductance in a finite quantum-anomalous-Hall
sample, and the two are consistent rather than in tension: the bulk
topological response is what carries no thermal noise, while the
quantised transport resides at the edge. The Corbino geometry is what
turns the bulk statement into a measurement, since it removes the edge
from the current path and leaves the bulk azimuthal Hall response as the
measured quantity; the prediction is then that this azimuthal noise
scales out the Chern number and reports the geometric sector alone.

\paragraph{Measurability of the excess noise.}
The geometric noise of Eq.~\eqref{eq:variance} is second order in the
drive and sits atop the equilibrium Johnson--Nyquist and shot noise, so
extracting it relies on its distinguishing signatures rather than its
magnitude: its $\mathcal{O}(E^2)$ bias dependence separates it from the
field-independent background, and its $D=f_0(1-f_0)$ temperature
envelope is a further independent handle. Because the topological
contribution cancels from it identically, the residual
$\mathcal{O}(E^2)$ transverse noise is a direct read-out of the
geometric variance $\Var_p(\Om_{\mathrm{geom}})$, which is the content
of the noise-spectroscopy proposal. Its absolute size is set by that
variance and the transport time $\tau$; a quantitative estimate for a
given polar metal is left to a dedicated experimental study.

\subsection{The null-space theorem is dimension-independent (3D)}

We now lift the noise analysis to three dimensions, where the driving
curvature carries all three sectors. The anomalous Hall current is the vector
$\bar J^i = -e^2\epsilon^{ijk}E_j (2\pi)^{-3}\!\int_{\BZ^3}\Om_k f\,\dd^3k$, and
its fluctuation couples to $\int_{\BZ^3}\Om_k\,\delta f$. Repeating the
derivation of Appendix~\ref{app:B} with the 3D correlator \eqref{eq:fdt_noise}, the
low-frequency current-noise tensor is governed by the covariance matrix of the
Berry-curvature vector under the thermal measure,
\begin{equation}\label{eq:covmatrix}
S^{ij}(0) \;\propto\; \epsilon^{iab}\epsilon^{jcd}E_a E_c\,
\mathcal{Z}\,\Cov_p(\Om_b,\Om_d),
\qquad
\Cov_p(\Om_b,\Om_d)=\langle\Om_b\Om_d\rangle_p-\langle\Om_b\rangle_p\langle\Om_d\rangle_p .
\end{equation}
Decomposing $\bm\Om=\langle\bm\Om\rangle+\bm\Om_\perp+\bm\Om_\parallel$ (harmonic,
exact, co-exact) and using that the harmonic representative is a \emph{constant}
vector, the harmonic contribution drops from Eq.~\eqref{eq:covmatrix} for
\emph{any} sampling measure $p$, since $\Cov_p(\text{const},\cdot)=0$. Physically
this is the same statement as in two dimensions: the constant harmonic flux
couples only to $\delta N\equiv0$. Hence:

\begin{theorem}[Dimension-independence of the harmonic null space]\label{thm:3d}
In any dimension the harmonic (topological) sector of the Berry curvature is a
constant form and therefore contributes exactly zero to the field-driven
current-noise covariance \eqref{eq:covmatrix}, for every chemical potential,
temperature, and band structure. The anomalous-Hall vector is a noiseless
quantum number in three dimensions as in two.
\end{theorem}

Theorem~\ref{thm:3d} needs no numerical corroboration, and we do not
present its numerical value as one: the harmonic block of $\Cov_p$ is the
covariance of a constant field and vanishes identically, by algebra rather than
by cancellation. In the lattice Weyl model \eqref{eq:weyl} it evaluates to the
floating-point roundoff floor, a relative $10^{-19}$--$10^{-17}$, at every grid
size, chemical potential, and temperature we have run, exactly as it must. The
two-way separation of topology from geometry therefore holds in three
dimensions. Note that this protects only the harmonic part of the topology; the
quantised slice-Chern jumps, which reside in the co-exact sector, are not
covered by Theorem~\ref{thm:3d} and are addressed next.

\subsection{The co-exact sector carries no clean noise signature}
The two-moment logic makes one cell of the dictionary conspicuous: the noise
signature of the co-exact (monopole) sector, which exists only in three
dimensions. The mean-transport chiral anomaly, Eqs.~\eqref{eq:proj}--\eqref{eq:quant},
sits cleanly in the co-exact sector because it couples to the \emph{divergence}
$\dd\Om$, which annihilates the other two sectors. One might hope the field-driven
\emph{noise} inherits an analogous clean co-exact channel. It does not, and the
reason is structural.

The field-driven current noise couples to $\bm\Om$ itself, not to its divergence,
through the covariance \eqref{eq:covmatrix}. Writing the exact and co-exact parts
as $\bm\Om_\perp$ and $\bm\Om_\parallel$, the noise covariance is
\begin{equation}\label{eq:threeway}
\Cov_p(\bm\Om) = \underbrace{\Cov_p(\bm\Om_\perp)}_{\text{exact}}
+ \underbrace{\Cov_p(\bm\Om_\parallel)}_{\text{co-exact}}
+ \underbrace{\Cov_p(\bm\Om_\perp,\bm\Om_\parallel)+\text{h.c.}}_{\text{cross}},
\end{equation}
the harmonic block having already dropped by Theorem~\ref{thm:3d}. The exact and
co-exact sectors are orthogonal in the \emph{uniform} $L^2$ inner product, but the
FDT samples them against the \emph{thermal} measure $p(\bm{k})=D(\bm{k})/\mathcal Z$,
which is peaked on the Fermi surface. Under a non-uniform measure orthogonality is
lost, and the cross term in Eq.~\eqref{eq:threeway} need not vanish. We find that
it does not.

\begin{figure}[t]
\centering
\includegraphics[width=\textwidth]{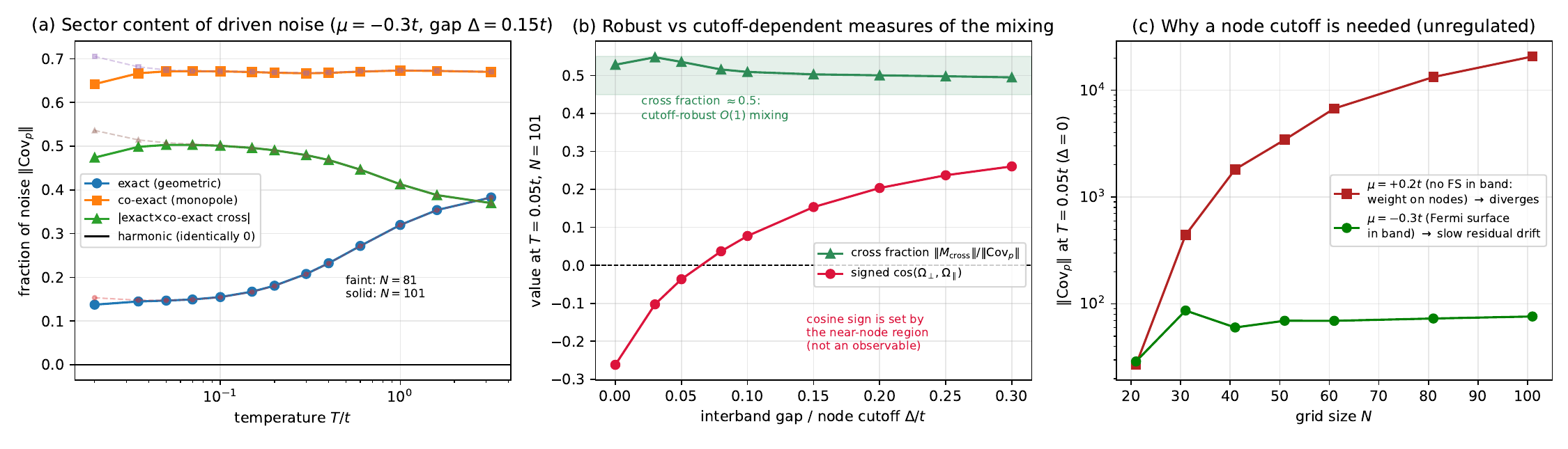}
\caption{\label{fig:noise}Sector structure of the field-driven current
noise in the lattice Weyl model, Eq.~\eqref{eq:weyl}, at metallic filling
$\mu=-0.3\,t$ (valence band; by particle-hole symmetry this equals the
conduction band at $\mu=+0.3\,t$).
(a)~Fraction of the noise-covariance norm $\|\Cov_p\|$ carried by each Hodge
sector versus temperature, at the physical interband gap $\Delta=0.15\,t$ that
renders the curvature square-integrable; solid symbols $N=101$, faint dashed
$N=81$ (they overlap, confirming convergence). The harmonic (topological) block
is identically zero (Theorem~\ref{thm:3d}); the fractions are norms of the
blocks of $\Cov_p=M_{\rm ex}+M_{\rm co}+M_{\rm cross}$ and need not sum to one,
the cross block being negative in the $zz$ channel. (b)~Robust versus
cutoff-dependent measures of the exact/co-exact mixing at $T=0.05\,t$, $N=101$,
as the node cutoff $\Delta$ is varied. The \emph{cross fraction}
$\|M_{\rm cross}\|/\|\Cov_p\|$ (green, shaded band $0.45$--$0.55$) is the
physical, cutoff-robust $\mathcal{O}(1)$ measure of non-separability; the
\emph{signed} overlap $\cos(\bm\Om_\perp,\bm\Om_\parallel)$ (red) is not an
observable, its sign is set by the near-node region and flips as the cutoff is
turned on, so it is quoted nowhere as a result. This is why the co-exact noise
is not separable (Proposition~\ref{thm:coexact}). (c)~Motivation for the
cutoff: the \emph{unregulated} ($\Delta=0$) $\|\Cov_p\|$ versus grid size $N$
at $T=0.05\,t$. With the Fermi surface inside the band ($\mu=-0.3\,t$) it drifts
only slowly (the exponentially suppressed residual node piece); with the thermal
weight pinned on the node singularities ($\mu=+0.2\,t$, above the valence-band
top) it grows without bound, the near-node curvature not being
square-integrable. Quantitative statements in the text are made only for the
regulated, grid-converged fractions.}
\end{figure}

Two features of the gapless band must be respected before quoting numbers.
First, the curvature diverges as $1/r^{2}$ at the nodes and is not
square-integrable, so $\Cov_p(\bm\Om)$ is ultraviolet-divergent whenever the
thermal weight at the nodes is nonzero. The divergence is regulated physically
by whatever cuts off the intraband semiclassics at the gapless points
(internode scattering; interband coherence once $|\varepsilon(\bm k)|\lesssim
T$) and is suppressed by the Boltzmann factor $e^{-|\mu|/T}$ at metallic
filling. Converged, cutoff-independent numbers therefore require the chemical
potential to sit a few $T$ inside a band; with the weight pinned on the nodes
the covariance grows without bound with grid size and is sensitive to the
node--grid commensuration (Fig.~\ref{fig:noise}(c)). Second, at electron doping
$\mu>0$ the Fermi surface lies in the conduction band, which then carries
essentially all of the thermal weight; by the particle-hole structure of
$H=\bm d\cdot\bm\sigma$ ($\varepsilon\to-\varepsilon$, $\bm\Om\to-\bm\Om$,
under which $\Cov_p$ is invariant) the conduction-band covariance at $+\mu$
equals the valence-band covariance at $-\mu$. We therefore work with the
valence band at $\mu=-0.3\,t$. Even here the exponentially small node weight
leaves a residual, slowly grid-growing piece: the \emph{absolute} covariance
$\|\Cov_p\|$ still drifts upward with $N$ (by $\sim\!10\%$ over
$N=61\text{--}101$) and the \emph{signed} exact/co-exact overlap is set by the
near-node region and does not converge. The physically meaningful content is
therefore carried not by the absolute norm but by the \emph{dimensionless}
block fractions of $\|\Cov_p\|$. To make every quantity separately
grid-convergent we impose the physical cutoff named above as an explicit
curvature regulator, an interband-coherence gap $\Delta$ (a single energy
scale, not to be confused with the node separation $\Delta k$ of
Eq.~\eqref{eq:bound}) acting as
$|\bm d|^{3}\!\to\!(|\bm d|^{2}+\Delta^{2})^{3/2}$, which bounds $\bm\Om$
(to $\sim\!1/\Delta^{2}$ at a node) and renders it square-integrable, so that
$\Cov_p$ and its Hodge blocks converge in $N$ for \emph{every} measure. Below
we quote fractions at a fixed $\Delta=0.15\,t$ (converged to $1\%$ by $N=101$)
and, crucially, verify that the conclusion is \emph{$\Delta$-robust}: the
quantity that carries the no-go is stable across the whole range
$0\le\Delta\le0.3\,t$, while the quantities that are not (the signed overlap,
the absolute norm) are flagged as such.

In this regulated regime the three geometric terms of Eq.~\eqref{eq:threeway}
carry comparable weight at physical temperatures. At $T=0.05\,t$ the Frobenius
fractions of $\|\Cov_p\|$ are $0.15$ (exact), $0.67$ (co-exact), and $0.50$
(cross), where the fractions are norms of the blocks of the matrix identity
$\Cov_p=M_{\rm ex}+M_{\rm co}+M_{\rm cross}$ and need not sum to one, the cross
block being \emph{negative} in some channels (Fig.~\ref{fig:noise}(a)). The
decisive quantity is the cross fraction. It is an $\mathcal{O}(1)$ mixing,
$\|M_{\rm cross}\|/\|\Cov_p\|\approx 0.5$, and, unlike the absolute norm or
the signed overlap, it is \emph{cutoff-robust}: it stays within $0.50$--$0.55$ (i.e.\ $0.52\pm0.03$)
across the entire regulator range $0\le\Delta\le0.3\,t$ and converges in the
grid to $1\%$ by $N=101$ (Fig.~\ref{fig:noise}(b)). The signed overlap
$\cos(\bm\Om_\perp,\bm\Om_\parallel)$ is by contrast \emph{not} an observable:
its magnitude is only $\mathcal{O}(0.1)$ and its sign is fixed entirely by the
near-node region, flipping from negative to positive as the node cutoff is
turned on (Fig.~\ref{fig:noise}(b)); no physical meaning attaches to its value,
and we quote it nowhere as a result. The mixing populates the same current
channels as the sector autocovariances [$\Cov_p=\mathrm{diag}(a,a,b)$ by the
$C_{4v}$ symmetry of the model]: in the $zz$ channel the co-exact
autocovariance, $11.2$, is more than twice the full covariance, $4.7$, the
excess being cancelled by the cross term, $-8.7$ (values at $N=101$,
$\Delta=0.15\,t$; the cross magnitude here \emph{exceeds} the full covariance
of the channel it sits in). No choice of drive direction or measured current
component therefore isolates the co-exact contribution; subtracting the exact
sector would require knowing it independently. The non-separability survives
symmetry breaking: an odd-in-$k_z$ polar distortion of the model [a $\sigma_z$
term that breaks the $k_z$ mirror and inversion; note this is \emph{not} a
Weyl-cone tilt $\propto\sigma_0$, which would leave $\bm\Om$ unchanged and
reshape only the thermal weight] leaves the cross fraction at $\approx0.5$,
indistinguishable from the symmetric model. (We caution that at $\Delta=0$ the
polar distortion \emph{appears} to strengthen the signed overlap; this is a
node--grid commensuration artefact, the distortion shifts the nodes off the
$k_0=\pi/2$ registry so the unregulated near-node samples fail to converge, and
it vanishes under the regulator, a further reason the signed overlap is not the
physical measure.) Raising the temperature offers no separation either: the
component-summed cross contribution does subside as the thermal measure
approaches uniformity, but the exact and co-exact autocovariances \emph{stay}
comparable and in the same channels ($0.23$ and $0.67$ of $\|\Cov_p\|$ at
$T=0.4\,t$), so the two geometric sectors remain entangled; the node
contribution that then dominates the measure is the cutoff-sensitive one, where
the intraband semiclassics is in any case uncontrolled; and the noise power
$\propto\mathcal{Z}\,\Cov_p$ of Eq.~\eqref{eq:covmatrix} offers no lever,
growing monotonically with $T$ and saturating, rather than collapsing, as the
measure approaches uniformity. We summarise this as:

\begin{lemma}[No conservation-law protection for the co-exact sector]\label{lem:coexact}
The co-exact field $\bm\Om_\parallel$ is an ordinary zero-mean fluctuating field:
it lies in the null space of no macroscopic collision invariant, so it enjoys
none of the measure-independent protection that silences the harmonic sector
(Theorem~\ref{thm:3d}). Consequently, for a generic non-uniform sampling measure
$p$ its cross-covariance with the exact sector,
$\Cov_p(\bm\Om_\perp,\bm\Om_\parallel)$, is not forced to vanish: the uniform-measure
Hodge orthogonality that guarantees $\langle\bm\Om_\perp,\bm\Om_\parallel\rangle_{L^2}=0$
constrains only the trace (component sum) of the cross-covariance matrix, not its
individual channels, and even the trace is lifted once $p$ is non-uniform.
\end{lemma}

\noindent Lemma~\ref{lem:coexact} is the structural obstruction; that the resulting
mixing is not merely nonzero but $\mathcal{O}(1)$, and cutoff-robustly so, is the
quantitative content, which we state as a proposition established by the regulated
numerics above:

\begin{proposition}[The co-exact mixing is $\mathcal{O}(1)$ and cutoff-robust]\label{thm:coexact}
For the lattice Weyl band at metallic filling with a non-degenerate thermal envelope
$p(\bm{k})=D(\bm{k})/\mathcal Z$, regulated by an interband gap $\Delta$, the cross
block of the field-driven current-noise covariance carries an $\mathcal{O}(1)$
fraction of $\|\Cov_p\|$, $\|M_{\rm cross}\|/\|\Cov_p\|\in[0.50,0.55]$ over
$0\le\Delta\le0.3\,t$, and populates the same current channels as the exact and
co-exact autocovariances. The co-exact sector is therefore not separable from the
exact sector by any choice of drive direction or measured current component. Raising
the temperature does not separate them: where the component-summed cross contribution
subsides, the exact and co-exact autocovariances remain comparable and in the same
channels, so no temperature achieves a separation.
\end{proposition}

The contrast with the harmonic sector is exact and instructive. The harmonic
sector is silenced by a \emph{conservation law} ($\delta N\equiv0$), which acts for
every measure and is therefore dimension-independent (Theorem~\ref{thm:3d}). The
co-exact sector enjoys no such law: it is an ordinary zero-mean fluctuating field,
and once the sampling measure is non-uniform it mixes with the exact sector like
any two non-orthogonal fields. The clean separation the noise provides is thus
two-way, topology (silent) versus geometry (noisy), and not three-way; the
monopole content, so sharply isolated in the mean by its divergence coupling
\eqref{eq:proj}, is not separately readable in the second moment. This fills the
last cell of the dictionary with a definite entry rather than an open question.

We note the one avenue this analysis does not close. The observable computed here
is the transverse (anomalous-Hall) current noise, which couples to $\bm\Om$. A
distinct observable, the fluctuation of the \emph{longitudinal} chiral-magnetic
($\bm E\parallel\bm B$) current, couples through the divergence source $\dd\Om$ and
might therefore inherit the clean co-exact projection of the mean response,
Eq.~\eqref{eq:proj}. Whether that longitudinal magnetoconductance noise resolves
the co-exact sector, or is instead dominated by internode scattering, is a
well-posed question left for future work; Proposition~\ref{thm:coexact} establishes
only that the transverse channel does not.

\section{Synthesis: the dictionary and noise spectroscopy}

The results assemble into a single sector-to-observable dictionary spanning both
moments of the current.

\begin{table}[t]
\renewcommand{\arraystretch}{1.35}
\centering
\begin{tabular}{@{}p{1.9cm} p{2.6cm} p{2.9cm} p{3.6cm} c@{}}
\hline
\textbf{Sector} & \textbf{Curvature role} & \textbf{Mean transport} & \textbf{Noise signature} & \textbf{Dim} \\
\hline
Harmonic & uniform topological flux & anomalous Hall (topological) & noiseless (FDT null space; proved dim.-independent) & 2D, 3D \\
Exact & Fermi-surface geometry & nonlinear-Hall Berry dipole & sole source of driven noise & 2D, 3D \\
Co-exact & Weyl-node / monopole source & chiral anomaly (node-quantised) & not separable in the transverse channel (Prop.~\ref{thm:coexact}) & 3D \\
\hline
\end{tabular}
\caption{\label{tab:dict} The Hodge dictionary for both moments of Berry-curvature
transport. The genuine synthesis is that one decomposition governs the first and
second moments, and that the noise separates topology from geometry because one
sector is silent and the other is the source.}
\end{table}

The experimental payoff is the separation principle. In the mean current
topology and geometry are superposed; in the noise they separate,
because the harmonic sector is silent
(Theorems~\ref{thm:fdt},~\ref{thm:3d}) and the exact sector is the
source. Current-noise spectroscopy therefore isolates local band
geometry from global band topology, in both two and three dimensions.
The dictionary is now closed on the noise side: the harmonic cell is a
proved null space in every dimension, the exact cell is the driven-noise
source, and the co-exact cell is a proved no-go for the transverse
channel, Proposition~\ref{thm:coexact}. The central claim is thus the
two-moment dictionary together with the two-way separation principle,
which is airtight; the monopole sector's noise is characterised, not
left implicit.

\section{The dimensional ladder and its endpoints}

The dimension column of the dictionary is the visible face of a ladder the
decomposition climbs:
\begin{center}
$d=3$: harmonic $+$ exact $+$ co-exact\qquad
$d=2$: harmonic $+$ exact\\
$d=1$: harmonic only (Zak phase $=$ polarisation)\qquad
$d=0$: no forms to decompose.
\end{center}
At $d=1$ the Brillouin zone is a circle, on which the Berry curvature
\emph{two}-form vanishes identically (there are no $2$-forms on a $1$-manifold),
so the object that survives is not the curvature but the Berry connection
$1$-form; its holonomy is the Zak phase, the modern theory of
polarisation~\cite{marzari2012}. In the Hodge decomposition of that $1$-form the
non-exact (harmonic) part is the gauge-invariant Zak phase, so the
one-dimensional Berry phase is purely harmonic in this precise sense. At $d=0$ the zone is a point and
the Berry phase relocates to the parameter manifold, its original home. Neither
endpoint is new physics, but they fix where the three-sector structure bottoms
out and make ``harmonic equals topological'' visible where nothing else survives.

The fluctuation side gives the low-dimensional end more than expository value.
Because the curvature vanishes in 1D, the exact-sector noise is identically zero
there, and the dictionary predicts that the Berry-sector contribution to a
one-dimensional polarisation response carries no thermal noise; what remains is
the quantum-metric (Bhowmick--Agarwal~\cite{BA26}) channel, so the two noise
mechanisms separate completely by dimension. A Thouless pump provides a
lower-risk realisation of the two-dimensional noise result on the $(k,t)$ torus:
the quantised pumped charge is harmonic and, by Theorem~\ref{thm:3d}, noiseless,
while the non-adiabatic and geometric corrections carry the exact-sector noise.
This exercises the clean two-way separation without invoking the co-exact sector,
which Proposition~\ref{thm:coexact} shows is not separable in any case.

\section{Conclusion}

The Hodge decomposition organises Berry-phase transport into three sectors whose
physical meanings are fixed by their mathematical type: harmonic forms carry
topology, exact forms carry Fermi-surface geometry, and co-exact forms carry the
monopole (chiral-anomaly) response. The same three sectors govern both moments of
the current. In the mean, each sector is a distinct, symmetry-protected transport
coefficient. In the noise, the harmonic sector is silenced by particle
conservation, a null-space protection we prove holds in every dimension
(Theorem~\ref{thm:3d}), while the exact sector is the sole source of field-driven
noise, so that current-noise spectroscopy separates global topology from local
geometry.

The one cell that the two-moment logic makes conspicuous, the noise signature
of the three-dimensional co-exact sector, we settle here: in the transverse
(anomalous-Hall) channel it is not a clean observable, because the co-exact sector
carries no conservation law and, under the thermal sampling measure, is strongly
non-orthogonal to the exact sector (Proposition~\ref{thm:coexact}). The clean
separation the framework delivers is therefore two-way, topology versus geometry,
and it holds in both two and three dimensions. This is the unified result: one
geometric decomposition that determines both moments of Berry-curvature transport,
with a measurable consequence in each and a proved boundary on how far the noise
separation extends.

\begin{acknowledgments}
The numerical results were obtained with the two-band lattice models described in
the text; the verification and figure-generation scripts are available with the
manuscript.
\end{acknowledgments}

\paragraph{Author contributions}
S.L.B.\ conceived the geometric framework and the fluctuation-dissipation
formulation; Z.-W.W.\ carried out the analytic and numerical calculations; both
authors wrote the manuscript.

\paragraph{Funding information}
The authors received no specific funding for this work.

\paragraph{Data and code availability}
The two-band lattice models are specified in the main text. The
verification and figure-generation scripts for the mean-transport
results and the scripts reproducing the three-dimensional fluctuation-dissipation
results of Sec.~4 are available upon request from the authors.

\appendix

\section{Positivity of the thermal weight and ellipticity}
\label{app:A}
The thermal weight is $D(\bm k)=f_0(1-f_0)=\tfrac14\,\mathrm{sech}^2[\beta(\varepsilon(\bm k)-\mu)/2]$.
On the compact Brillouin torus the band energy $\varepsilon(\bm k)$ is continuous,
hence bounded: $|\varepsilon(\bm k)-\mu|\le M<\infty$ for some finite $M$ (for the
lattice model~\eqref{eq:weyl}, $M$ is a few $t$). Since $\mathrm{sech}^2$ is
monotone decreasing in its squared argument,
\begin{equation}
D(\bm k)\;\ge\;\tfrac14\,\mathrm{sech}^2\!\big(\beta M/2\big)\;=:\;D_{\min}(T)\;>\;0
\qquad\text{for every }\bm k\text{ and every }T>0 .
\end{equation}
Thus $D$ is strictly bounded away from zero on the whole zone at any positive
temperature (it degenerates only in the strict limit $T\to0$, where the envelope
collapses onto the Fermi surface). The diffusion matrix of the Fokker--Planck
operator associated with Eq.~\eqref{eq:fdt_noise} is, on the subspace orthogonal
to the single conserved mode $\int\delta f=0$, proportional to $D(\bm k)$ acting
by multiplication; the lower bound $D\ge D_{\min}>0$ makes it uniformly elliptic
on that physical subspace. The negative rank-one term $-D(\bm k)D(\bm k')/\mathcal Z$
in Eq.~\eqref{eq:fdt_noise} is exactly the projector that removes the zero
eigenvalue associated with total particle number, and no other; the remaining
spectrum is bounded below by $D_{\min}$ times the spectral gap of the bare
exchange kernel.

\section{The excess-noise formula \texorpdfstring{\eqref{eq:variance}}{}}
\label{app:B}
We derive Eq.~\eqref{eq:variance} from the particle-conserving
Langevin--Boltzmann equation. Linearising about equilibrium and Fourier
transforming in time, the fluctuation $\delta f_{\bm k}(t)$ under a
relaxation-time collision operator obeys the Ornstein--Uhlenbeck dynamics
$\partial_t\delta f_{\bm k}=-\delta f_{\bm k}/\tau+\xi_{\bm k}(t)$, with the noise
correlator~\eqref{eq:fdt_noise}. The Hall-current fluctuation, after the harmonic
term has been removed by particle conservation (Theorem~\ref{thm:fdt}), is a
linear functional
\begin{equation}
\delta J(t)=A\!\int_{\BZ}a(\bm k)\,\delta f_{\bm k}(t)\,\dd^d k,
\qquad a(\bm k)=\Om_{\mathrm{geom}}(\bm k),\qquad
A=-\frac{e^2E_x}{(2\pi)^d}.
\end{equation}
The stationary equal-time covariance of the Ornstein--Uhlenbeck process is\\
$\langle\delta f_{\bm k}\,\delta f_{\bm k'}\rangle
=\tfrac{\tau}{2}\langle\xi_{\bm k}\xi_{\bm k'}\rangle/\delta(t-t')
=\frac{(2\pi)^d}{V_{\mathrm{sys}}}\big[D_{\bm k}\delta(\bm k-\bm k')
-D_{\bm k}D_{\bm k'}/\mathcal Z\big]$,
and the two-time correlator decays as $e^{-|t-t'|/\tau}$. The zero-frequency
current noise is therefore
\begin{align}
S(0)&=\int_{-\infty}^{\infty}\!\dd t\,\langle\delta J(t)\delta J(0)\rangle
   =A^2\,(2\tau)\!\iint a_{\bm k}a_{\bm k'}
     \frac{(2\pi)^d}{V_{\mathrm{sys}}}
     \Big[D_{\bm k}\delta(\bm k-\bm k')-\frac{D_{\bm k}D_{\bm k'}}{\mathcal Z}\Big]
     \dd^d k\,\dd^d k' \nonumber\\
  &=A^2\,\frac{2\tau(2\pi)^d}{V_{\mathrm{sys}}}
     \Big[\!\int a^2 D\,\dd^d k-\frac{1}{\mathcal Z}\Big(\!\int a D\,\dd^d k\Big)^{\!2}\Big]
   =A^2\,\frac{2\tau(2\pi)^d}{V_{\mathrm{sys}}}\,
     \mathcal Z\,\Var_p(a),
\end{align}
where $\int t\,e^{-|t|/\tau}$-type integration gives the factor $2\tau$, and in
the last step\\ $\int a^2D-\mathcal Z^{-1}(\int aD)^2
=\mathcal Z[\langle a^2\rangle_p-\langle a\rangle_p^2]=\mathcal Z\Var_p(a)$ with
$p=D/\mathcal Z$. Inserting $A^2=(e^2E_x/(2\pi)^d)^2$ and $d=2$ reproduces
Eq.~\eqref{eq:variance}. The derivation is unchanged in three dimensions
component by component, giving the covariance-tensor form~\eqref{eq:covmatrix}
with $a\to\Om_b$; the harmonic (constant) part of $a$ contributes zero to
$\Var_p$ and to $\Cov_p$, which is Theorems~\ref{thm:fdt} and~\ref{thm:3d}.


\begin{thebibliography}{99}

\bibitem{haldane2004}
F.~D.~M. Haldane, \textit{Berry Curvature on the Fermi Surface: Anomalous Hall Effect as a Topological Fermi-Liquid Property}, Phys. Rev. Lett. \textbf{93}, 206602 (2004), \doi{10.1103/PhysRevLett.93.206602}.

\bibitem{tknn1982}
D.~J. Thouless, M.~Kohmoto, M.~P. Nightingale, and M.~den Nijs, \textit{Quantized Hall Conductance in a Two-Dimensional Periodic Potential}, Phys. Rev. Lett. \textbf{49}, 405 (1982), \doi{10.1103/PhysRevLett.49.405}.

\bibitem{sodemann2015}
I.~Sodemann and L.~Fu, \textit{Quantum Nonlinear Hall Effect Induced by Berry Curvature Dipole in Time-Reversal Invariant Materials}, Phys. Rev. Lett. \textbf{115}, 216806 (2015), \doi{10.1103/PhysRevLett.115.216806}.

\bibitem{ma2019}
Q.~Ma \emph{et al.}, \textit{Observation of the nonlinear Hall effect under time-reversal-symmetric conditions}, Nature \textbf{565}, 337 (2019), \doi{10.1038/s41586-018-0807-6}.

\bibitem{sundaram1999}
G.~Sundaram and Q.~Niu, \textit{Wave-packet dynamics in slowly perturbed crystals: Gradient corrections and Berry-phase effects}, Phys. Rev. B \textbf{59}, 14915 (1999), \doi{10.1103/PhysRevB.59.14915}.

\bibitem{xiao2010}
D.~Xiao, M.-C. Chang, and Q.~Niu, \textit{Berry phase effects on electronic properties}, Rev. Mod. Phys. \textbf{82}, 1959 (2010), \doi{10.1103/RevModPhys.82.1959}.

\bibitem{BA26}
D.~Bhowmick and A.~Agarwal, \textit{Berry connection polarizability induced nonlinear thermal noise}, Phys. Rev. B \textbf{113}, 235405 (2026), \doi{10.1103/PhysRevB.113.235405}.

\bibitem{NCM13}
T.~Neupert, C.~Chamon, and C.~Mudry, \textit{Measuring the quantum geometry of Bloch bands with current noise}, Phys. Rev. B \textbf{87}, 245103 (2013), \doi{10.1103/PhysRevB.87.245103}.

\bibitem{nakahara2003}
M.~Nakahara, \textit{Geometry, Topology and Physics}, CRC Press, Boca Raton, FL (2003).

\bibitem{frankel2011}
T.~Frankel, \textit{The Geometry of Physics: An Introduction}, Cambridge University Press (2011), \doi{10.1017/CBO9781139061377}.

\bibitem{fhs2005}
T.~Fukui, Y.~Hatsugai, and H.~Suzuki, \textit{Chern Numbers in Discretized Brillouin Zone: Efficient Method of Computing (Spin) Hall Conductances}, J. Phys. Soc. Jpn. \textbf{74}, 1674 (2005), \doi{10.1143/JPSJ.74.1674}.

\bibitem{armitage2018}
N.~P. Armitage, E.~J. Mele, and A.~Vishwanath, \textit{Weyl and Dirac semimetals in three-dimensional solids}, Rev. Mod. Phys. \textbf{90}, 015001 (2018), \doi{10.1103/RevModPhys.90.015001}.

\bibitem{fukushima2008}
K.~Fukushima, D.~E. Kharzeev, and H.~J. Warringa, \textit{The Chiral Magnetic Effect}, Phys. Rev. D \textbf{78}, 074033 (2008), \doi{10.1103/PhysRevD.78.074033}.

\bibitem{zhong2016}
S.~Zhong, J.~E. Moore, and I.~Souza, \textit{Gyrotropic magnetic effect and the magnetic moment on the Fermi surface}, Phys. Rev. Lett. \textbf{116}, 077201 (2016), \doi{10.1103/PhysRevLett.116.077201}.

\bibitem{mapesin2015}
J.~Ma and D.~A. Pesin, \textit{Chiral magnetic effect and natural optical activity in metals with or without Weyl points}, Phys. Rev. B \textbf{92}, 235205 (2015), \doi{10.1103/PhysRevB.92.235205}.

\bibitem{kumar2021}
D.~Kumar \emph{et al.}, \textit{Room-temperature nonlinear Hall effect and wireless radiofrequency rectification in Weyl semimetal TaIrTe$_4$}, Nat. Nanotechnol. \textbf{16}, 421 (2021), \doi{10.1038/s41565-020-00839-3}.

\bibitem{nye1985}
J.~F.~Nye, \textit{Physical Properties of Crystals}, Oxford University Press, Oxford (1985).

\bibitem{marzari2012}
N.~Marzari, A.~A. Mostofi, J.~R. Yates, I.~Souza, and D.~Vanderbilt, \textit{Maximally localized Wannier functions: Theory and applications}, Rev. Mod. Phys. \textbf{84}, 1419 (2012), \doi{10.1103/RevModPhys.84.1419}.

\bibitem{Kubo66}
R.~Kubo, \textit{The fluctuation-dissipation theorem}, Rep. Prog. Phys. \textbf{29}, 255 (1966), \doi{10.1088/0034-4885/29/1/306}.

\bibitem{LL57}
L.~D.~Landau and E.~M.~Lifshitz, \textit{On hydrodynamic fluctuations}, Sov. Phys. JETP \textbf{5}, 512 (1957).

\bibitem{Braun26}
S.~L.~Braunstein, \textit{Physical completion of the Navier-Stokes equations}, arXiv:2605.21357 (2026).

\bibitem{GG15}
A.~J.~Bestwick, E.~J.~Fox, X.~Kou, L.~Pan, K.~L.~Wang, and D.~Goldhaber-Gordon, \textit{Precise Quantization of the Anomalous Hall Effect near Zero Magnetic Field}, Phys. Rev. Lett. \textbf{114}, 187201 (2015), \doi{10.1103/PhysRevLett.114.187201}.

\end{thebibliography}
\end{document}